\documentclass[aps,prl,twocolumn,nofootinbib]{revtex4-1}

\usepackage{graphicx}
\usepackage{dcolumn}
\usepackage{bm}
\usepackage{ textcomp }
\usepackage{color}
\usepackage{amsmath}
\usepackage{amssymb}
\usepackage{xcolor}
\usepackage{hyperref}
\usepackage{easyReview}
\usepackage{mathdots}
\usepackage{float}
\usepackage{lineno}
\usepackage{array}
\usepackage{xcolor,colortbl}
\usepackage[normalem]{ulem}
\definecolor{LightBlue}{rgb}{0.8,0.8,0.8}

\usepackage{amsmath} 
\newcommand{\angstrom}{\text{\normalfont\AA}}

\begin{document}
\title{Thermal Hall Effect in the Kitaev-Heisenberg System with Spin-Phonon Coupling}

\date{\today}

\author{Shaozhi Li}
\email{lis1@ornl.gov}
\author{Satoshi Okamoto}
\affiliation{Materials Science and Technology Division, Oak Ridge National Laboratory, Oak Ridge, TN 37831, USA}
\begin{abstract}
 We investigate the thermal Hall effect in a Kitaev-Heisenberg system, a model for Kitaev candidate $\alpha$-RuCl$_3$, in the presence of the coupling between spin and phonon arising from chlorine atoms' vibration. 
  We observe that the coupling modifies the relative stability between different magnetic states under a magnetic field, especially stabilizing a canted zigzag antiferromagnetic state.
  Remarkably, the spin-phonon interaction has distinct effects on the thermal Hall conductivity in different magnetically ordered states. For the canted zigzag state, which is relevant to $\alpha\text{-RuCl}_3$, the spin-phonon interaction enhances the energy gap that is induced by a magnetic field and suppresses the thermal Hall conductivity at low temperatures.
  Importantly, we find that the spin-phonon interaction destabilizes the quantized thermal Hall effect in the spin liquid state.
  Our results demonstrate a crucial role of phonon degrees of freedom to the thermal Hall effect in Kitaev materials. 
\end{abstract}

\maketitle


{\it Introduction.} Phonons play an essential role in shaping electronic and magnetic properties in materials~\cite{WeberPRL1987,LazzeriPRB2008,MarchandPRL2010,NowadnickPRL2012,ShaozhiPRB2017,SousPRL2018,Costa2020,CostaPRL2021,LeePRB2021,SapkotaPRB2021,Shaozhinpj,XingPRL2021}. For example, the electron-phonon interaction induces the charge-density-wave and superconductivity in the high-temperature superconductor $\text{Ba}_{0.51}\text{K}_{0.49}\text{BiO}_3$~\cite{WenPRL2018}. Besides, phonons can couple to spins and induce the spin-Peierls transition in quasi-one-dimensional materials, including 
$\text{CuGeO}_3$~\cite{HasePRL1993} and  TiOCl~\cite{ShazPRB2005}.

Recently, a significant interplay between phonons and spins was observed in the Kitaev material $\alpha\text{-RuCl}_3$~\cite{PlumbPRB2014,KubotaPRB2015,SandilandsPRL2015,JohnsonPRB2015,SearsPRB2015,CaoPRB2016,KoitzschPRL2016,YamadaPRL2017,Kasahara2018nature,HentrichPRL2018,HaoxiangNC2021,SaiMu,Blair2022}.
In $\alpha\text{-RuCl}_3$, the edge-sharing Ru-Cl octahedra form a honeycomb lattice with an effective spin-$1/2$. The interaction between two nearest-neighbor pseudospins strongly depends on the Ru-Ru distance and the Ru-Cl-Ru bond angle~\cite{KimPRB2016}. 
In $\alpha\text{-RuCl}_3$, both the spin and phonon behaviors are modified by the spin-phonon interaction. For example, the spin-phonon coupling renormalizes phonon propagators and generates the salient Fano lineshape in the Raman spectroscopy~\cite{GaominPRM2019,GlamazdaPRB,Kexin,Metavitsiadis,SandilandsPRL2015}. Besides, the coupling of phonons to Majorana fermions plays a pivotal role in realizing the quantized thermal Hall effect in $\alpha\text{-RuCl}_3$~\cite{MengxingPRL2018,VinklerPRX2018,MetavPRB2020,MengxingPRR2020,FengPRB2021,RuizPRB2021}.

To gain insight into the novel magnetic behavior in $\alpha\text{-RuCl}_3$, various theoretical models were developed by using the density-functional theory (DFT) or fitting to inelastic neutron scattering spectra~\cite{PlumbPRB2014,KimPRB2015,JohnsonPRB2015,WinterPRB2016,SANDILANDSprb2016,HouPRB2017,WangPRB2017,CookmeyerPRB2018,SuzukiPRB2018,EichstaedtPRB2019,Laurellnpj2020}. Although the spin-phonon interaction is not included,
these models generate a number of experimentally related predictions, including the heat capacity and the Raman spectroscopy. However, some important experimental features remained unexplained, such as the Fano effect in the Raman spectroscopy~\cite{Kexin,Metavitsiadis} and the sample dependence of the quantized thermal Hall effect in $\alpha\text{-RuCl}_3$~\cite{YamashitaPRB2020}.
It has been proposed that the spin-phonon coupling is essential to resolve these difficulties. In this case, it is necessary to develop a model that includes both spin and phonon degrees of freedoms. However, compared to the Kitaev-Heisenberg model, far fewer studies exist for such a model. Recently, several studies have considered phonons in the Majoranan fermion representation to explain the Fano effect~\cite{FengPRB2021,Kexin,Metavitsiadis}. Such studies only considered the Kitaev-type interaction and neglected some significant spin-spin interactions in $\alpha\text{-RuCl}_3$. Furthermore, the effect of the spin-phonon coupling on the thermal Hall conductivity has not been clarified, although some interesting predictions have been made by analyzing a phenomenology model~\cite{YePRL2018}.

\begin{figure}[t]
\center\includegraphics[width=0.95\columnwidth]{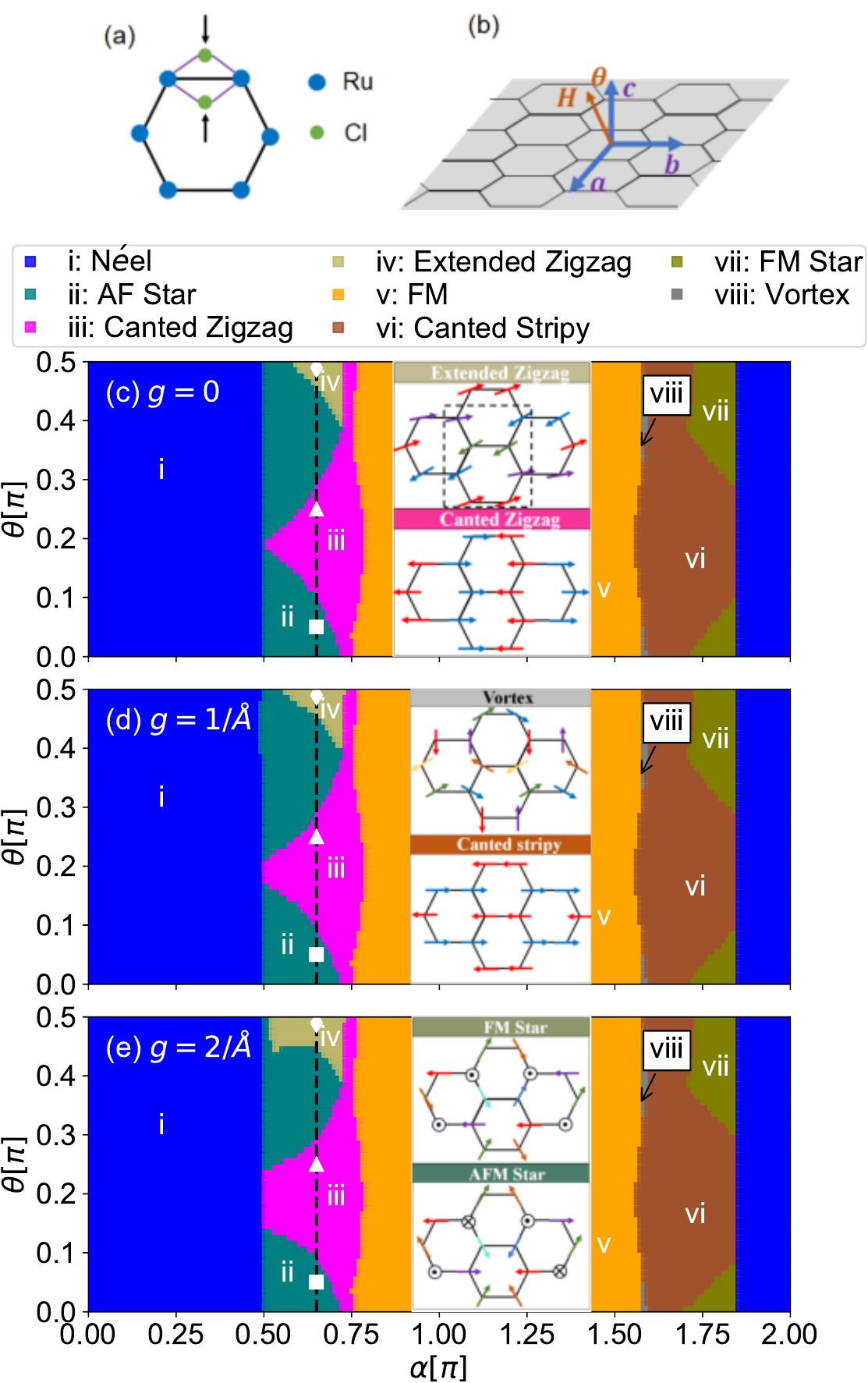}
\caption{\label{Fig:fig2} (a) A sketch of chlorine atoms' vibration in $\alpha\text{-RuCl}_3$. (b) A sketch of the Honeycomb lattice that lies on the $ab$ plane. Panels (c)-(d) plot the phase diagrams for the spin-phonon interaction $g=0$, $1/\AA$, and $2/\AA$ under a magnetic field $g_L\mu_B|{\bf H}|=0.4A$, respectively. The insets in panels (c)-(e) show the spin configuration of each magnetic state when the magnetic field is along the $c$ axis. The spin configuration is projected onto the $ab$ plane.}
\end{figure}

In this work, we develop a microscopic model with a spin-phonon interaction to study the thermal Hall conductivity in $\alpha\text{-RuCl}_3$. We consider the optical phonon, which originates from the vibration of the Cl-Cl bond (shown in Fig.~\ref{Fig:fig2}(a)). This optical phonon modifies the Ru-Cl-Ru bond angle, hereby the spin-spin interaction between Ru atoms~\cite{KimPRB2016,KaibPRB2021}. We first use the mean-field theory to study the spin-phonon interaction in the Kitaev-Heisenberg model under a magnetic field. We observe that optical phonons stabilize the canted zigzag state and destabilize the antiferromagnetic star state. 
Next, we use the generalized spin-wave theory to study the thermal Hall conductivity. We find that the spin-phonon interaction has distinct effects on the thermal Hall conductivity in different magnetic states. For example, in the canted zigzag state, which is relevant to $\alpha$-RuCl$_3$, the spin-phonon interaction increases the energy gap and suppresses the thermal Hall conductivity at low temperatures. 
In other magnetic states, i.e. the antiferromagnetic star state, the spin-phonon interaction generates a nonmonotonic behavior in the thermal Hall conductivity.
Importantly, by examining the pure Kitaev model coupled with phonons, we observe that the quantized thermal Hall effect in the spin liquid state is destabilized by the spin-phonon interaction. 
Our results provide {\it qualitative} guidance to understand the thermal Hall conductivity in $\alpha\text{-RuCl}_3$.

{\it Model Hamiltonian.} The optical mode of the chlorine atom's motion in $\alpha\text{-RuCl}_3$ is shown in Fig.~\ref{Fig:fig2} (a). This optical phonon changes the distance $d$ between two chlorine atoms, resulting in modifying the spin-spin interaction $J(d)$ between Ru atoms. Assuming the variation of $d$ is small, the interaction can be approximated as $J(d)=J(d_0)\left[1 - g(d-d_0)\right]$, where $d_0$ is the distance at the equilibrium position, and the spin-phonon coupling $g=-\frac{1}{J(d_0)}\frac{\partial J(d)}{\partial d}$. In this case, the proposed extended-Kitaev-Heisenberg model~\cite{chernnpj2021} for $\alpha\text{-RuCl}_3$ may be modified as $H=H_{\mathrm{spin}} + H_{\mathrm{ph}} + H_{\mathrm{spin-ph}}$,
where 
\begin{eqnarray}
H_{\mathrm{spin}}&=&\sum_{\langle ij \rangle_\gamma} \left[ 2K\hat{S}_i^{\gamma} + J \hat{{\bf S}}_i \cdot \hat{{\bf S}}_j +\right] -\sum_{i,\alpha} g_L\mu_B H_\alpha \hat{S}_i^{\alpha},\nonumber\\
H_{\mathrm{ph}}&=&\sum_{\langle ij \rangle} \left[\frac{M\hat{\dot{u}}_{ij}^2}{2} + \frac{K\hat{u}_{ij}^2}{2} \right],\nonumber\\
H_{\mathrm{spin-ph}}&=&-\sum_{\langle ij\rangle_{\gamma}} g\hat{u}_{ij} \left[ 2K\hat{S}_i^{\gamma} \hat{S}_j^{\gamma} + J\hat{{\bf S}}_i \cdot \hat{{\bf S}}_j   \right]. \nonumber 
\end{eqnarray}
Here, $J$ and $K$ are the nearest-neighbor Heisenberg and Kitaev couplings. $\langle ij \rangle_{\gamma}$ denotes the nearest-neighbor $\gamma$ bond, with $\gamma=x,y,z$. $(\alpha,\beta,\gamma)=(y,z,x), (z,x,y)$, and $(x,y,z)$ for the $x$, $y$, and $z$ bonds, respectively. We follow Ref.~\cite{KoyamaPRB2021} and set that the [111] direction in the spin space is parallel to the $c$ axis and $S^z$ direction is on the $ac$ plane. 
$H_{\alpha}$ is the magnetic field along the $\alpha$ direction, which is shown in Fig.~\ref{Fig:fig2}(b). The magnetic field strength is labeled as $|{\bf H}|=\sqrt{H_a^2+H_b^2+H_c^2}$.  $\hat{u}_{ij}$ denotes the variation of the distance between two chlorine atoms, and $\langle \hat{u}_{ij} \rangle = d_{ij} - d_0$. $M$ and $K$ are the mass of the chlorine atom and the elastic constant between chlorine atoms, respectively. Here, we assume that spin-phonon couplings for the Kitaev and Heisenberg interactions are the same. This assumption does not change our conclusions on the effect of the spin-phonon interaction in the canted zigzag state (see details in the supplementary material~\cite{KaibPRB2021,Supple}).

The ground state is obtained using the mean-field approach~\cite{KoyamaPRB2021,Supple}. Specifically, we set $\hat{S}^\gamma = \langle S^\gamma \rangle + \delta\hat{S}^{\gamma}$ and
$\hat{u} = \langle u \rangle + \delta\hat{u}$. The Hamiltonian is then rewritten as $H=\mathrm{const} + H^{(1)} + H^{(2)} + H^{(3)}$, where $H^{(1)}$, $H^{(2)}$, and $H^{(3)}$ include terms with one, two, and three operators, respectively. 
$\langle S^\gamma \rangle$ and $\langle u \rangle$ are obtained by 
self-consistently solving $H^{(1)}$
(see supplementary material~\cite{Supple}). 
After obtaining stable spin and lattice configurations, we apply a generalized spin-wave theory to $H^{(2)}$ and compute the Berry curvature~\cite{Supple}.
The thermal Hall conductivity is calculated via $\kappa_{xy}=-\frac{k_B^2T}{\hbar N}\sum_{k}^{BZ}\sum_{n=1}^L c_2(f_{\mathrm{BE}}(\epsilon_{nk}))\Omega_{nk}$~\cite{KatsuraPRL2010,MatsumotoPRB2014,YangPRL2020},
where $c_2(x)$ is given by $c_2(x)=\int_0^x dt \left(\mathrm{ln}\frac{1+t}{t}  \right)^2$, $f_{\mathrm{BE}}(\epsilon_{nk})$ is the Bose distribution function, and BZ denotes the Brillouin zone. 
$\Omega_{nk}$ is the Berry curvature at the momentum $k$ for the $n$th band.
$2L$ is the number of bands, and $N$ is the number of the magnetic unit cell, which is labeled as a dashed-rectangular in the inset of Fig.~\ref{Fig:fig2}.
In this work, we set $N=120\times 120$, $\hbar=k_B=1$ and neglect three-particle interaction $H^{(3)}$.

\begin{figure}[t]
\center\includegraphics[width=0.95\columnwidth]{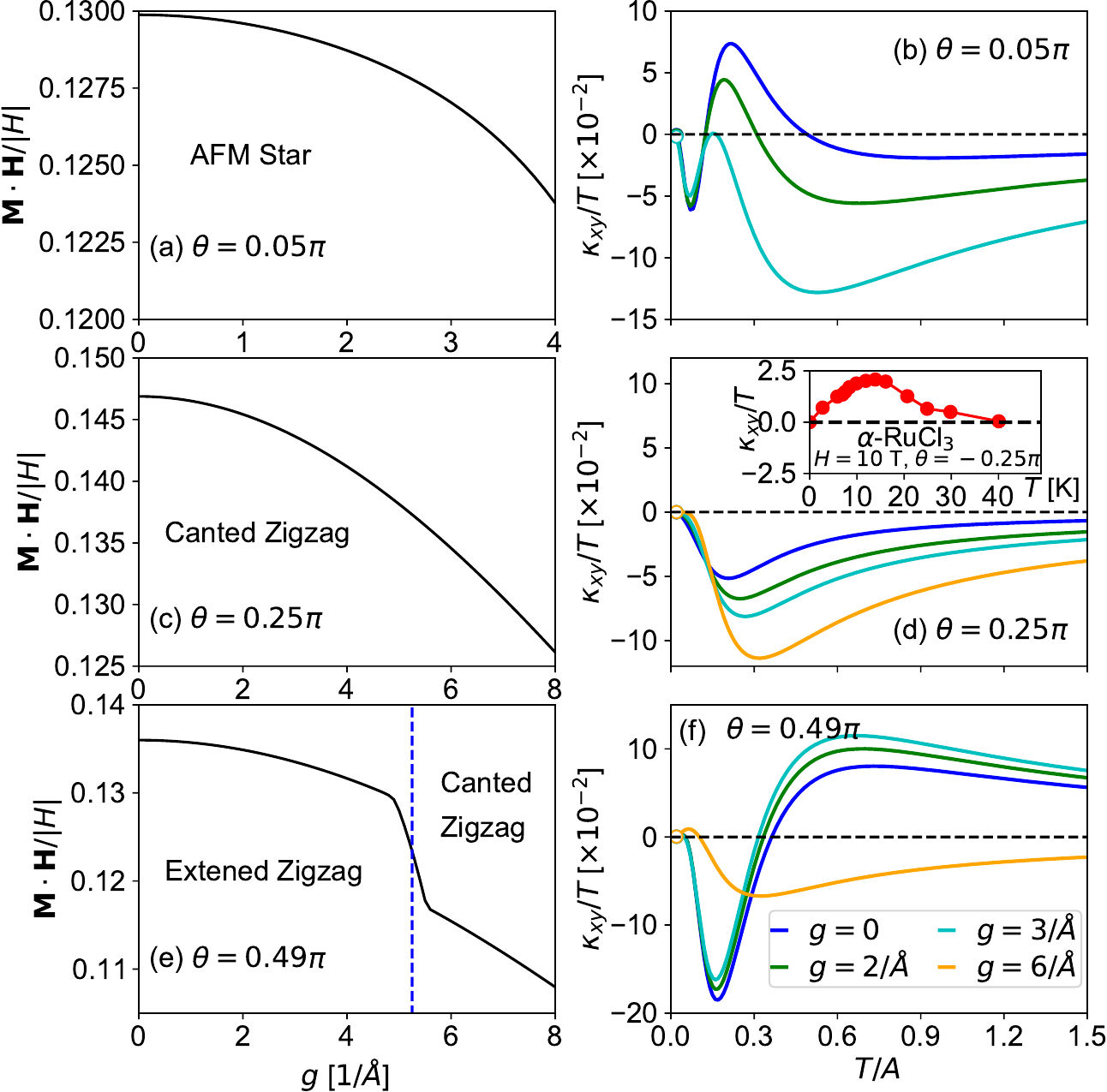}
\caption{\label{Fig:fig3} The left column shows the magnetization ${\bf M}\cdot{\bf H}/|{\bf H}|$ along the field direction for three different values of $\theta$. The blue dashed line in panel (c) represents the phase boundary between the extended zigzag  and canted zigzag states. The right column shows the corresponding thermal Hall conductivity $\kappa_{xy}/T$ as a function of temperature $T$. The inset in panel (e) plots the thermal Hall conductivity of $\alpha\text{-RuCl}_3$, obtained from Ref.~\cite{Yokoiscience2021}. Here, we set $\alpha=0.65\pi$ and $g_L\mu_B|{\bf H}|=0.4 A$. }
\end{figure}

{\it Results.} Considering a recent experiment on $\alpha\text{-RuCl}_3$~\cite{HaoxiangNC2021}, we set the phonon frequency $\omega_{\mathrm{ph}}=8$ meV, hereby the elastic constant $K=535.2$ meV/\angstrom$^2$. Here, we set $K={A}\mathrm{sin}\alpha$, $J={A}\mathrm{cos}\alpha$~\cite{JanssenPRL2016,ChernPRB2017,ConsoliPRB2020,KoyamaPRB2021}. Ref.~\cite{HaoxiangNC2021} reported that the optical phonon energy is close to the spin-spin interaction in $\alpha\text{-RuCl}_3$ at low temperatures, hereby we set $A=\omega_\text{ph}$. The results for other parameters are shown in the supplementary material~\cite{Supple}.  At zero field, the ground state of the classical model is the N\'{e}el state for $-0.15\pi<\alpha<0.5\pi$, the zigzag state for $0.5\pi<\alpha<0.85\pi$, the ferromagnetic (FM) state for $0.85\pi<\alpha<1.5\pi$, and the stripy state for $1.5\pi<\alpha<1.85\pi$~\cite{JanssenPRL2016,Supple}. These ground states are changed under a magnetic field~\cite{JanssenPRL2016,KoyamaPRB2021}. In this work, we rotate the magnetic field in the $ac$ plane (shown in Fig.~\ref{Fig:fig2}(b)) and label the angle between the magnetic field and the $c$ axis as $\theta$. Figure~\ref{Fig:fig2}(c) shows the phase diagram in the plane of $\alpha$ and $\theta$ at $g_L\mu_B|{\bf H}|=0.4 A$ and $g=0$.
Besides these four magnetic states, there appear another four magnetic states, including anti-ferromagnetic star state (AF Star), extended zigzag state, ferromagnetic star state (FM Star), and vortex state (vortex)~\cite{JanssenPRL2016,KoyamaPRB2021,Supple}. Remarkably, the spin-phonon interaction changes the  magnetic ground state and modifies the phase diagram. As shown in Figs.~\ref{Fig:fig2}(d) and~\ref{Fig:fig2}(e), the spin-phonon interaction increases the regions of the canted zigzag and extended zigzag states by suppressing the AF Star state when $0.5\pi<\alpha<0.75\pi$. By further increasing the spin-phonon interaction, the extended zigzag state is replaced by the canted zigzag state (see Fig.~\ref{Fig:fig3}). These results imply that optical phonons stabilize the zigzag state under a magnetic field.

We now analyze the effect of the spin-phonon interaction on the magnetization ${\bf M}$ and the thermal Hall conductivity $\kappa_{xy}$ in the parameter region of the zigzag state that exists at zero field. If not stated otherwise, we set $\alpha=0.65\pi$ and $g_L\mu_B|{\bf H}|=0.4 A$ in this work. The results for other parameters are shown in the supplementary material~\cite{Supple}. The value $\alpha=0.65\pi$ is labeled by the dashed line in Figs.~\ref{Fig:fig2} (c) -~\ref{Fig:fig2} (e).
Figures~\ref{Fig:fig3}(a) -~\ref{Fig:fig3}(c) plot the magnetization for $\theta=0.05\pi$, $0.25\pi$, and $0.49\pi$, respectively. The ground states for these three values of $\theta$ are the AFM Star, canted zigzag, and extended zigzag states, labeled as a white square, triangle, and diamond in Figs.~\ref{Fig:fig2} (c) -~\ref{Fig:fig2}(e), respectively. Figures~\ref{Fig:fig3}(a) -~\ref{Fig:fig3}(c) show that the spin-phonon interaction suppresses the magnetization for all these three states. In addition, a first-order phase transition from the extended zigzag state to the canted zigzag state occurs at $\theta=0.49\pi$ and $g=5.6/\r{A}$, accompanied by a sudden drop in ${\bf M}\cdot{\bf H}/|{\bf H}|$.

\begin{figure}[t]
\center\includegraphics[width=\columnwidth]{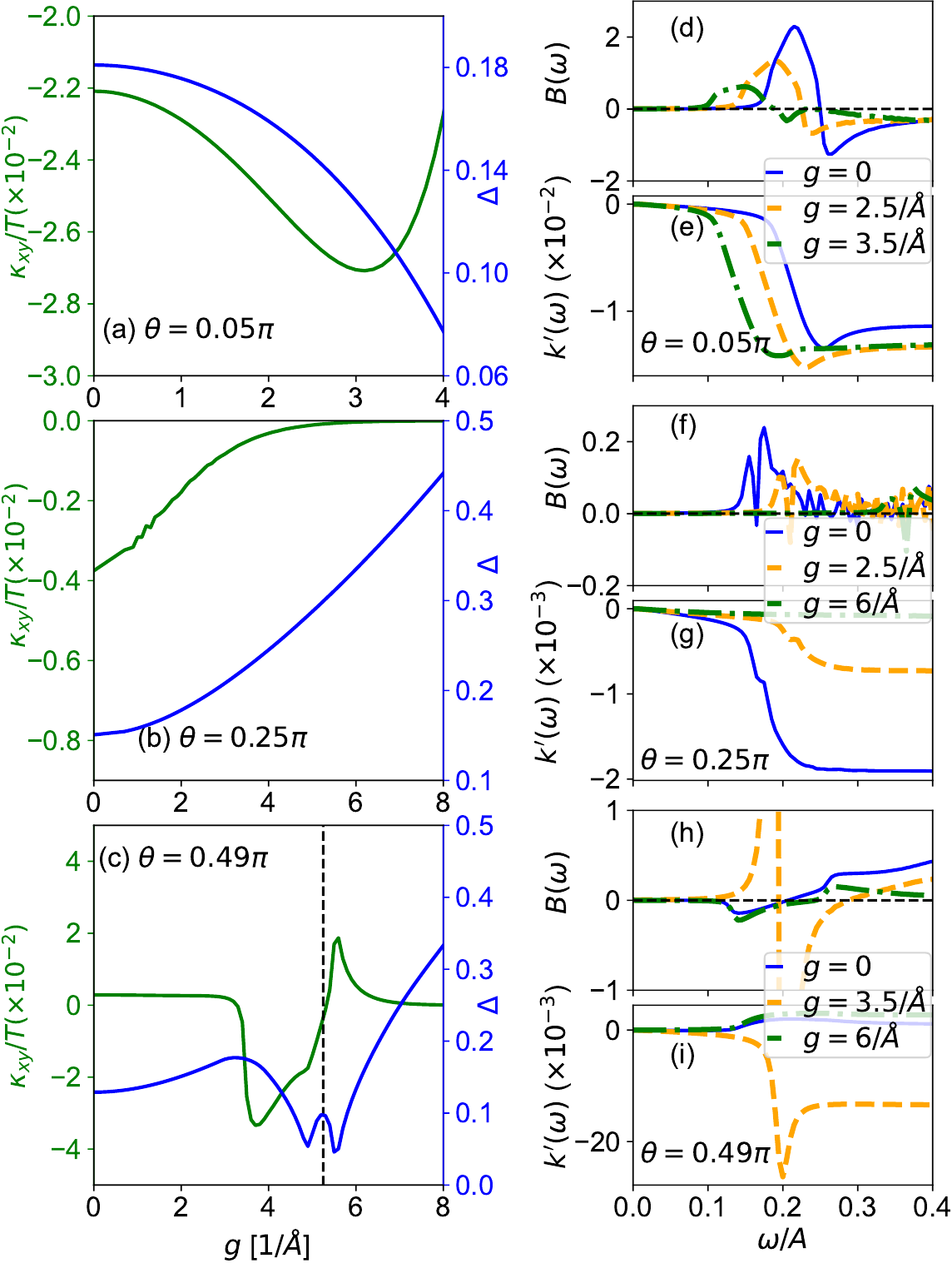}
\caption{\label{Fig:fig4} The left column plots the thermal Hall conductivity $\kappa_{xy}/T$ (green curve) and the energy gap $\Delta$ (blue curve) as a function of the spin-phonon interaction $g$ for three different values of $\theta$. The black dashed line in panel (c) represents the phase boundary between the extended zigzag state and the canted zigzag state. Panels (d), (f), and (h) plot the Berry curvature density $B(\omega)$, and panels (e), (g), and (i) plot the integrated Berry curvature density $K^\prime(\omega)$. Here, we set $\alpha=0.65\pi$, $T=0.04 A$, and $g_L\mu_B|{\bf H}|=0.4 A$.}
\end{figure}

Figures~\ref{Fig:fig3}(d) -~\ref{Fig:fig3}(f) plot the thermal Hall conductivity $\kappa_{xy}/T$ as a function of temperature $T$ for several different values of $g$. Interestingly, $\kappa_{xy}/T$ has distinct temperature dependences in different magnetic states. We examine $\kappa_{xy}/T$ at various values of $\alpha$ and find that $\kappa_{xy}/T$ has three and two extrema in the AFM Star and extended zigzag staes, respectively. $\kappa_{xy}/T$ could have either one or two extrema in the canted zigzag state, depending on the value of $\alpha$. While our predicted $\kappa_{xy}/T$ is much smaller than that in $\alpha\text{-RuCl}_3$ (see supplementary material~\cite{Supple}), we note that the temperature-dependent $\kappa_{xy}/T$ in the canted zigzag state is qualitatively consistent with that measured in $\alpha\text{-RuCl}_3$~\cite{Yokoiscience2021} (see the inset of Fig.~\ref{Fig:fig3}(e)). Although it would be interesting to study the microscopic mechanism for this similarity~\cite{YangPRL2020}, such a study is beyond the scope of our work. $\kappa_{xy}/T$ in the intermediate temperature regime ($1.2\mathrm{ A}>T>0.3$ A) has been significantly changed by the spin-phonon interaction for all these three states. 

To better visualize the change in $\kappa_{xy}/T$ at low temperatures, we plot $\kappa_{xy}/T$ in Figs.~\ref{Fig:fig4}(a) -~\ref{Fig:fig4}(c) for $T=0.04A$. The energy gap $\Delta$ (blue curve) is also plotted in these three panels.
At $\theta=0.05\pi$ (AF Star), $|\kappa_{xy}|/T$ is enhanced by the spin-phonon coupling when $g<3/\angstrom$ and suppressed when $g>3/\angstrom$. This nonmonotonic behavior
arises from the competition between the changes in $\Delta$ and $\Omega_{n,k}$. 
The latter contribution can be visualized more clearly by the Berry curvature density, which is defined as 
$B(\omega)=\sum_{n,k}\Omega_{n,k}\delta(\omega-\epsilon_{n,k})$ and plotted in Figs.~\ref{Fig:fig4}(d),~\ref{Fig:fig4}(f), and~\ref{Fig:fig4}(h).
The $\delta$ function is evaluated using the Lorentzian function with a width of $0.005$. 
In the AF Star state, the spin-phonon interaction monotonically decreases $\Delta$, leading to an enhancement of $|\kappa_{xy}|/T$.
However, the Berry curvature density near $\omega=0.2A$ is suppressed by the spin-phonon interaction, resulting in the decrease in $|\kappa_{xy}|/T$. Figure~\ref{Fig:fig4} (e) plots the integrated Berry curvature density $K^\prime(\omega)=-\int_0^{\omega} c_2(f_{\rm BE}(\omega^\prime))B(\omega^\prime)d\omega^\prime$, equivalent to the thermal Hall conductivity. It is easy to observe this competition from $K^\prime(\omega)$ in Fig.~\ref{Fig:fig4}(e). At $\theta=0.25\pi$ (canted zigzag), $|\kappa_{xy}|/T$ is suppressed by the spin-phonon interaction because $\Delta$ increases and $B(\omega\approx0.2A)$ decreases with $g$.

At $\theta=0.49\pi$, $\kappa_{xy}/T$ is positive and slightly suppressed by $g$ when $g<3/\angstrom$. This behavior is attributed to the enhanced energy gap. Near $g=3.2/\angstrom$, $\kappa_{xy}/T$ suddenly becomes negative and increases with $g$. This sudden change is due to the band crossing at the $\Gamma$ point (see details in the supplementary material~\cite{Supple}). 
By further increasing $g$, $\Delta^\prime$ increases. At $g=5.6/\angstrom$, the phase transition occurs, and the ground state becomes the canted zigzag state. $|\kappa_{xy}|/T$ in this canted zigzag state has the same behavior as that shown in Fig.~\ref{Fig:fig4} (b).

\begin{figure}[t]
\center\includegraphics[width=\columnwidth]{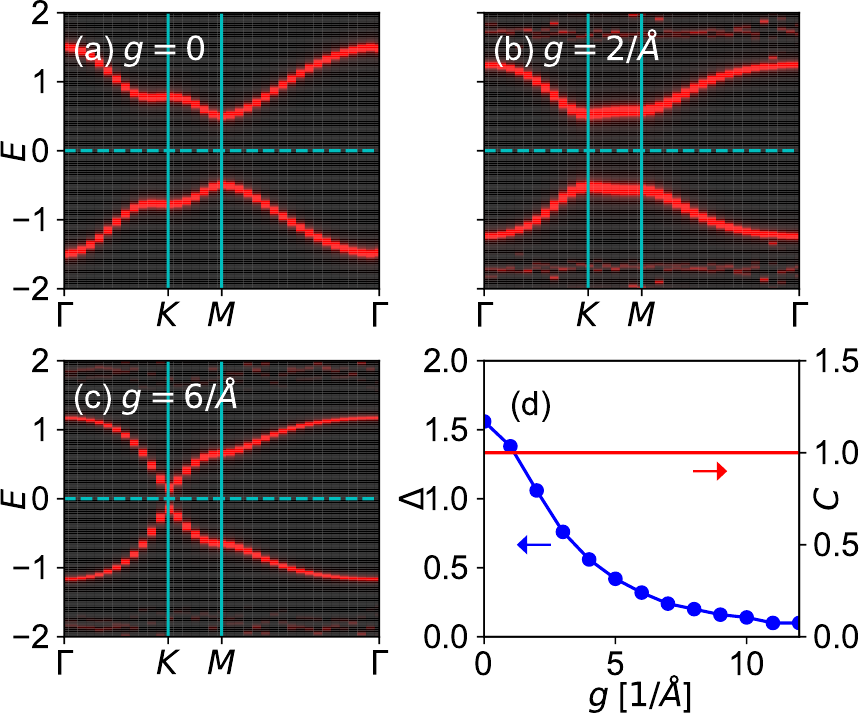}
\caption{\label{Fig:fig5} Panels (a)-(c) plot the spectral functions of the Kitaev-phonon model for three different spin-phonon coupling strengths $g$. Panel (d) plots the energy difference $\Delta$ (blue curve) between the energies of the lowest excited states above and below the Fermi surface at the $K$ point. The red line represents the Chern number $C$.}
\end{figure}

Finally, we turn our attention to the spin-phonon interaction in the spin liquid state. We consider the pure Kitaev model coupled with phonons, which is given by $H=\frac{i}{2}\sum_{\langle ij \rangle} K(1-g\hat{u}_{ij}) c_ic_j+\frac{i\tilde{h}}{2}\sum_{\langle\langle ij \rangle\rangle}c_ic_j+H_\text{ph}$~\cite{Yamada2020}. Here, $c_i$ is the Majorana fermion operator at site $i$, and $\tilde{h}$ is an effective magnetic field. We set $K=1$, $\tilde{h}=0.3$, and phonon frequency $\Omega=1$ in our calculations and use the second-order perturbation theory to study this model (see details in the supplementary material~\cite{Supple}). Figure~\ref{Fig:fig5} plots the spectral functions for three different values of $g$. There exist energy gaps at zero energy due to an applied magnetic field. In the low-energy region, the spin-phonon interaction decreases the energy $E_{\pm}(k)$ of the lowest excited state at the $K$ point. Here, $E_{+(-)}(k)$ represents the energy above (below) the Fermi surface at the momentum $k$. Figure~\ref{Fig:fig5} (d) plots the variation of the energy difference $\Delta$ between $E_{+}(K)$ and $E_{-}(K)$. While $\Delta$ is decreased by the spin-phonon interaction, the Chern number $C$ remains 1~\cite{Fukui,VanhalaPRL2016}. Thus, the spin-phonon interaction does not induce a topological transition.

The nonzero Chern number implies a quantized thermal Hall conductivity at low temperatures. However, when the thermal energy is close to or larger than the energy gap, the quantized thermal Hall effect disappears~\cite{MirmojarabianPRB2020}. With increasing the spin-phonon interaction, the energy gap approaches zero, hereby the highest temperature that can observe the quantized thermal Hall effect is reduced to lower. Therefore, the sample-dependence of the half-integer quantized thermal Hall effect in $\alpha$-RuCl$_3$ could be ascribed to the sample-dependent spin-phonon interactions~\cite{YamashitaPRB2020}.


{\it Summary.} In this work, we studied the effect of the spin-phonon interaction on the magnetic ground state and the thermal Hall conductivity in the Kitaev-Heisenberg system. We observed that a large spin-phonon interaction stabilizes the canted zigzag state under a magnetic field. Further, 
the spin-phonon interaction has distinct effects on the thermal Hall conductivity in different magnetic states. In the zigzag state, the spin-phonon interaction increases the energy gap and suppresses the low temperature thermal Hall conductivity. This effect is independent of parameters in the canted zigzag state for the Kitaev-Heisenberg model~\cite{Supple}. Importantly, we find that the spin-phonon interaction destabilizes the quantized thermal Hall effect in the spin liquid state.
Our results provide crucial guidance to understand thermal Hall conductivity in Kitaev materials, including $\alpha\text{-RuCl}_3$ and $\text{Na}_2\text{Co}_2\text{TeO}_6$~\cite{HongPRB2021}.

In this work, we consider optical phonons arising from the displacement of chlorine atoms. It would be important to study acoustic phonons, which involve the displacement of Ruthenium atoms, and examine how these phonons modify the magnetic properties in $\alpha\text{-RuCl}_3$.  

\begin{acknowledgments}
This research was supported by the U.S. Department of Energy, Office of Science, National Quantum Information Science Research Centers, Quantum Science Center. 
This research used resources of the Compute and Data Environment for Science (CADES) at the Oak Ridge National Laboratory, which is supported by the Office of Science of the U.S. Department of Energy under Contract No. DE-AC05-00OR22725. 
\end{acknowledgments}

\bibliography{main}

\end{document}